\documentclass[prl,aps,twocolumn,showpacs, floatfix]{revtex4}
\usepackage{amssymb}
\usepackage{hyperref}
\usepackage{graphicx}
\usepackage{dcolumn}
\usepackage{bm,amsmath,verbatim}
\usepackage{mathrsfs}

\def\be{\begin{equation}}
\def\ee{\end{equation}}
\def\bea{\begin{eqnarray}}
\def\eea{\end{eqnarray}}
\def\bse{\begin{subequations}}
\def\ese{\end{subequations}}

\def\be{\begin{eqnarray}}
\def\ee{\end{eqnarray}}

\begin{document}

\title{Anisotropic Weyl Fermions from Quasiparticle Excitation Spectrum of a
3D Fulde-Ferrell Superfluid}
\author{Yong Xu$^{1}$}
\author{Rui-Lin Chu$^{1}$}
\author{Chuanwei Zhang$^{1}$}
\thanks{Corresponding Author, Email: chuanwei.zhang@utdallas.edu}
\affiliation{$^{1}$Department of Physics, The University of Texas at Dallas, Richardson,
Texas 75080, USA}

\begin{abstract}
Weyl fermions, first proposed for describing massless chiral Dirac fermions
in particle physics, have not been observed yet in experiments. Recently,
much effort has been devoted to explore Weyl fermions around band touching
points of single particle energy dispersions in certain solid state
materials (named \textit{Weyl semimetals}), similar as graphene for Dirac
fermions. Here we show that such Weyl semimetals also exist in the
quasiparticle excitation spectrum of a three-dimensional (3D) spin-orbit
coupled Fulde-Ferrell (FF) superfluid. By varying Zeeman fields, the
properties of Weyl fermions, such as their creation and annihilation, number
and position, as well as anisotropic linear dispersions around band touching
points, can be tuned. We study the manifestation of anisotropic Weyl
fermions in sound speeds of FF fermionic superfluids, which are detectable
in experiments.
\end{abstract}

\maketitle


\emph{Introduction} Weyl fermions \cite{Weyl} are massless chiral Dirac
fermions with linear energy dispersions in momentum space. Recently, the
existence of Weyl fermions has been explored in various solid state
materials, such as Pyrochlore Iridates~\cite{Wan2011prb,Aji2012prb},
ferromagnetic compound HgCr$_{2}$Se$_{4}$~\cite{ZhongFang2011prl},
multilayer topological insulators~\cite{Burkov2011PRL}, photonic crystals~%
\cite{LingLu2013NP}, as well as in optical lattices~\cite%
{Bercioux2009PRA,Lan2012PRB}. These materials, named as Weyl semimetals,
possess band touching points in their single particle energy spectrum,
around which the energy dispersions are linear and can be described by
chiral Weyl equation. These band touching points (i.e., Weyl nodes) appear
in pairs with opposite topological invariance~\cite{volovik}. In contrast to
two dimensional (2D) Dirac fermions (e.g., graphene), which are unstable
against perturbations that break time-reversal or spatial inversion
symmetries, Weyl nodes are stable and the only way to destroy them is to
merge two Weyl nodes with opposite topological invariances.

The recent progress in experimental observations of Majorana fermions (half
of a regular Dirac fermion) using quasiparticle excitations in solid state
topological superconductors~\cite{MourikScience2012,DengMTNano2012,
DasNature2012,LPRnature2012} leads to a nature question: whether Weyl
fermions can also be observed in the quasiparticle excitation spectrum
(instead of single particle spectrum) of superconductors or superfluids
(e.g., $^{3}$He A phase)~\cite{volovik,Gong2011prl,Sumanta2013PRA}. While
Majorana fermions in superconductors emerge as quasiparticle excitations in
real space (inside defects) and low dimensions (1D or 2D), Weyl fermions
describe energy dispersions in momentum space in 3D. Therefore the
semiconductor/superconductor heterostructures for observing Majorana
fermions, where superconductivity is induced through proximity effects, are
not suitable for the observation of Weyl fermions.

The recent experimental realization of spin-orbit (SO) coupling~\cite%
{Lin2011Nature,Jing2012PRL,
Zwierlen2012PRL,PanJian2012PRL,Qu2013PRA,Spilman2013PRL} in ultracold atomic
gases provide another platform for exploring a variety of intriguing
physics, including topological superfluids with Majorana fermions \cite%
{Zhang2008PRL,Sato2009PRL,ShiLiang2011PRL,LJiang2011PRL,XJLiu2012PRA,Melo2012PRL,Gong2012PRL, XiongXiv2013}%
. In particular, the low-temperature phase diagram of spin-orbit coupled
Fermi gases is dominated by Fulde-Ferrell (FF) superfluids with finite
momentum pairing~\cite{FuldePR1964,Larkin1964,
Zheng2013PRA,FanWu2013PRL,Liu2013PRA,Fu2013PRA,LinDongArx,Hui2013PRA,Iskin2013,YongArx13,XJliu13PRA}
in the presence of an in-plane Zeeman field, even in 3D. In 1D and 2D, such
FF superfluids can support Majorana fermions \cite%
{Qu2013NC,Yi2013NC,XJ2013PRA,Chun2013PRL}. However, whether such FF
superfluid state can support Weyl fermion excitations has not been explored.

In this Letter, we show that Weyl fermions can emerge from quasiparticle
excitation spectrum of a 3D FF superfluid. The system we consider is a 3D
degenerate Fermi gas with Rashba spin-orbit coupling (in $xy$ plane) and
Zeeman fields (in-plane ($h_{x}$) and out-of-plane ($h_{z}$)). The in-plane
Zeeman field breaks the spatial inversion symmetry of the Fermi surface,
yielding finite momentum pairing \cite{Zheng2013PRA}. The rich phase
diagrams in such 3D FF superfluids are obtained. In suitable parameter
regions, we find band touching points between particle and hole branches in
the quasiparticle excitation spectrum of the FF superfluid, which possess
non-zero topological invariances and anisotropic linear dispersions along
all three directions, indicating the existence of anisotropic Weyl fermion
excitations. The properties of Weyl fermions, including their number and
position, creation and annihilation, and anisotropy, can be controlled by
varying Zeeman fields and interaction strength between atoms. Finally, we
investigate the signature of anisotropic Weyl fermion excitations in the
speeds of sound of the FF fermionic superfluids, which are measurable in
experiments.

\begin{figure}[t]
\includegraphics[width=2.8in]{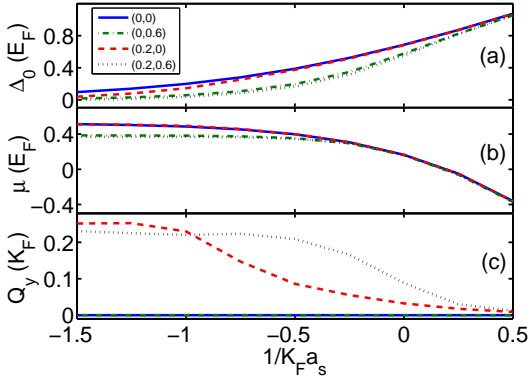}
\caption{(Color online) Plot of $\Delta _{0}$ in (a), $\protect\mu $ in (b),
and $Q_{y}$ in (c) as a function of $1/K_{F}a_{s}$ for different parameters (%
$h_{x}$,$h_{z}$). $\protect\alpha K_{F}=E_{F}$ and the temperature $T=0$.}
\label{cross}
\end{figure}

\emph{Model and effective Hamiltonian}: We consider a 3D Fermi gas with
\textit{s}-wave contact interactions. The many-body Hamiltonian can be
written as $H=\int d\mathbf{r}\hat{\Psi}^{\dagger }(\mathbf{r})H_{s}(\hat{%
\mathbf{p}})\hat{\Psi}(\mathbf{r})-U\int d\mathbf{r}\hat{\Psi}_{\uparrow
}^{\dagger }(\mathbf{r})\hat{\Psi}_{\downarrow }^{\dagger }(\mathbf{r})\hat{%
\Psi}_{\downarrow }(\mathbf{r})\hat{\Psi}_{\uparrow }(\mathbf{r})$, where
the single particle Hamiltonian $H_{s}(\hat{\mathbf{p}})=\frac{\hat{\mathbf{p%
}}^{2}}{2m}-\mu +H_{\text{SOC}}(\hat{\mathbf{p}})+H_{z}$ with momentum
operator $\hat{\mathbf{p}}=-i\hbar \nabla $, chemical potential $\mu $,
attractive interaction strength $U$, and the atom mass $m$; the Rashba SO
coupling $H_{\text{SOC}}(\hat{\mathbf{p}})=\alpha (\hat{\mathbf{p}}\times
\mathbf{\sigma })\cdot {\mathbf{e}_{z}}$ with Pauli matrix $\mathbf{\sigma }$%
; the Zeeman field is along $x$ (in-plane) and $z$ (out-of-plane)
directions, $H_{z}=h_{x}\sigma _{x}+h_{z}\sigma _{z}$. $\hat{\Psi}(\mathbf{r}%
)=[\hat{\Psi}_{\uparrow }(\mathbf{r}),\hat{\Psi}_{\downarrow }(\mathbf{r}%
)]^{T}$ and $\hat{\Psi}_{\nu }^{\dagger }(\mathbf{r})$ ($\hat{\Psi}_{\nu }(%
\mathbf{r})$) is fermionic atom creation (annihilation) operator.

The thermodynamical potential in mean-field approximation can be written as
\begin{eqnarray}
\Omega &=&|\Delta |^{2}/U+\sum\nolimits_{\mathbf{k}}\left( \hbar ^{2}(-%
\mathbf{k}+\mathbf{Q}/2)^{2}/2m-\mu \right) \\
&&-\sum\nolimits_{\mathbf{k},\sigma }\frac{1}{2\beta }\ln (1+e^{-\beta E_{%
\mathbf{k}\sigma }}).  \notag
\end{eqnarray}%
Here $E_{\mathbf{k}\sigma }$ is the eigenvalue of $4\times 4$ Bogoliubov-de
Gennes (BdG) Hamiltonian
\begin{equation}
H_{B}=\left(
\begin{array}{cc}
H_{s}(\mathbf{k}+\mathbf{Q}/2) & \Delta _{0} \\
\Delta _{0} & -\sigma _{y}H_{s}(-\mathbf{k}+\mathbf{Q}/2)^{\ast }\sigma _{y}%
\end{array}%
\right) ,
\end{equation}%
$\mathbf{Q}=Q_{y}\mathbf{e}_{y}$ is the total momentum of the Cooper pair
induced by the deformation of the Fermi surface~\cite{Zheng2013PRA}. The
mean-field solutions of $\Delta _{0}$, $Q_{y}$, and $\mu $ satisfy the
saddle point equations $\partial \Omega /\partial \Delta _{0}=0$, $\partial
\Omega /\partial Q_{y}=0$, and the atom number equation $\partial \Omega
/\partial \mu =-n$ with a fixed total atom density $n$. To regularize the
ultra-violet divergence at large $\mathbf{k}$, we follow the standard
procedure~\cite{StringariRMP} $\frac{1}{U}=\frac{m}{4\pi \hbar ^{2}a_{s}}%
-\int \frac{d\mathbf{k}}{(2\pi )^{3}}\frac{m}{\hbar ^{2}k^{2}}$ with the
s-wave scattering length $a_{s}$. The self-consistent solution is obtained
through the minimization of the free energy $F=\Omega +\mu n$. The energy
unit is chosen as the Fermi energy $E_{F}=\hbar ^{2}\mathbf{K}_{F}^{2}/2m$
of non-interacting Fermi gases without SO coupling and Zeeman fields with
Fermi vector $K_{F}=(3\pi ^{2}n)^{1/3}$.

\begin{figure}[t]
\includegraphics[width=3.4in]{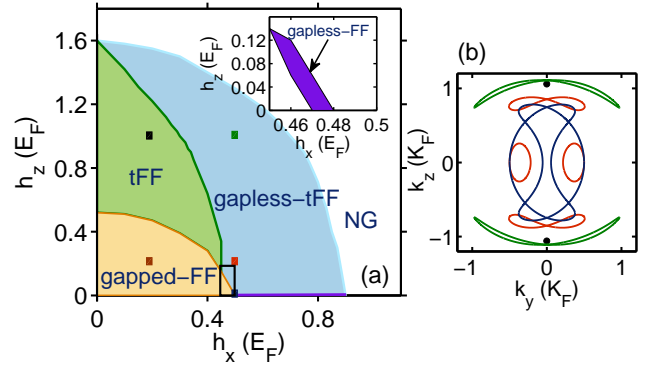}
\caption{(a) Mean-field phase diagrams of 3D spin-orbit coupled Fermi gases.
The area in the small black box is enlarged in the inset, which shows the
gapless FF phase. NG: Normal Gas. (b) Contours of zero energy quasiparticle
spectrum in the plane $(k_{y},k_{z})$ with $k_{x}=0$. Here $h_{x}=0.5E_{F}$,
$h_{z}=0$ (blue line), $h_{z}=0.2E_{F}$ (red line), $h_{z}=E_{F}$ (green
line), and $h_{x}=0.2E_{F}$, $h_{z}=E_{F}$ (black points) correspond to
blue, red, green, black square points in (a). The brown point in the gapped
FF phase has no zero energy excitations. In both figures, $\protect\alpha %
K_{F}=E_{F}$ and $1/a_{s}K_{F}=-0.1$.}
\label{phase}
\end{figure}

\emph{BCS-BEC crossover and} \emph{phase diagram}: In Fig.~\ref{cross}, we
plot the change of $\mu $, $\Delta _{0}$ and $Q_{y}$ in the BCS-BEC
crossover at zero temperature. With increasing $1/K_{F}a_{s}$, $\Delta _{0}$
increases and $\mu $ decreases, signalling the crossover from BCS
superfluids to BEC molecules. In the BEC limit, $\Delta _{0}$, $\mu $ and $%
Q_{y}$ become the same for different values of $(h_x,h_z)$ since fermion
atoms form bound molecules. Henceforth we focus on the BCS region. In this
region, $\Delta _{0}$ is smaller when the total Zeeman field $h=\sqrt{%
h_{x}^{2}+h_{z}^{2}}$ is larger. The nonzero $Q_{y}$ in the presence of $%
h_{x}$ indicates the existence of Cooper pairings with finite center-of-mass
momenta.

In Fig.~\ref{phase}(a), we map out the zero-temperature mean-field phase
diagram in the $(h_{x},h_{z})$ plane. In 3D Fermi gases, quantum
fluctuations generally do not change the phase diagram qualitatively~\cite%
{Lian2013PRA,Devreese2013Arxiv}. The uniform superfluid phase (with zero
total momentum pairing), which exists at $h_{x}=0$, is replaced by the
gapped FF phase~\cite{Hui2013PRA,LinDongArx} in the presence of a small
in-plane Zeeman field $h_{x}$ because of the broken inversion symmetry of
the Fermi surface. To characterize different phases, we consider the
quasiparticle gap $E_{g}$ that represents the energy difference between the
minimum of the particle branch and the maximum of the hole branch ($E_{g}>0$%
, gapped; $E_{g}\leq 0$, gapless). The topological FF (tFF) phase that
supports Weyl fermions in the quasiparticle excitation spectrum appears in
the large $h_{z}$ region. The topological FF phase is gapped except at Weyl
nodes, where $E_{g}=0$. The critical $h_{z}$ for the transition to the tFF
phase decreases as $h_{x}$ increases and reaches the minimum $%
h_{z}=0.14E_{F} $ at $h_{x}=0.45E_{F}$. In the absence of $h_{z}$, the
gapless FF phase appears with a large range of $h_{x}$~\cite%
{Hui2013PRA,LinDongArx}, while the region is extremely small (shown in the
inset of Fig.~\ref{phase}(a)) in the presence of $h_{z}$. Instead, a new
phase called gapless topological FF phase dominates the remaining parameter
region because of the large total Zeeman field. This phase possesses
non-topological gapless excitations in additional to topological Weyl
fermions, which is different from the topological FF phase where Weyl nodes
are the only gapless excitations. This phase should also be distinguished
from the gapless FF state because its band structure has certain topological
property.

\begin{figure}[t]
\includegraphics[width=3.4in]{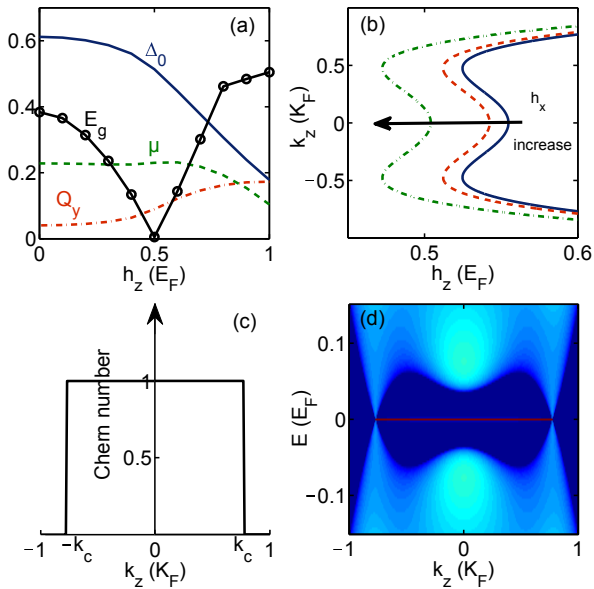}
\caption{(Color online) (a) The change of the quasiparticle excitation gap $%
E_{g}$, $\Delta _{0}$, $Q_{y}$, $\protect\mu $ as a function of $h_{z}$ with
$h_{x}=0.2E_{F}$. (b) Gap closing points ($E_{g}=0$) in the $(h_{z},k_{z})$
plane for $h_{x}=0$ (solid blue line), $h_{x}=0.1$ (dashed red line), and $%
h_{x}=0.2$ (dashed-dot green line). (c) Chern number for a fixed $k_{z}$
plane. (d) displays the density of states, which is calculated via the
iterative Green function method~\protect\cite{XiDai2008PRB}. $k_{y}=0$. The
light blue region and the red line (Fermi arch) represent bulk and surface
excitations, respectively. In (c) and (d), $h_{x}=0.2E_{F}$, $%
h_{z}=0.55E_{F} $. In all four figures $\protect\alpha K_{F}=E_{F}$, $%
1/a_{s}K_{F}=-0.1$.}
\label{TS_property}
\end{figure}

\begin{figure}[t]
\includegraphics[width=3.4in]{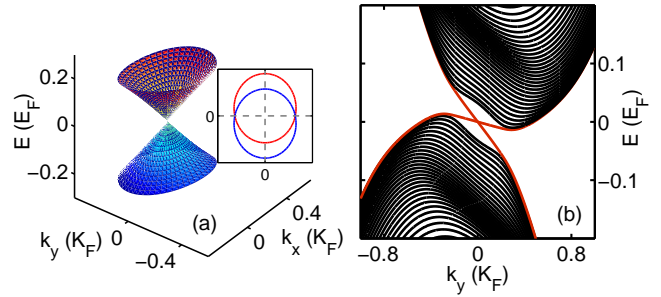}
\caption{(Color online) (a) Quasiparticle excitations around the Weyl node $%
\mathbf{k}_{W}=\left( 0,0,0.77K_{F}\right) $ with $h_{z}=0.55E_{F}$ and $%
h_{x}=0.2E_{F}$. The inset gives the contours of energy with $E=0.1E_{F}$ (
red line) and $E=-0.1E_{F}$ (blue line). (b) Quasiparticle excitation
spectrum in the gapless topological FF phase as a function of $k_{y}$ with
fixed $k_{z}=0.8K_{F}$ and confinement in the $x$ direction. Black lines are
bulk states while the red lines correspond to the surface state. $%
h_{x}=0.5E_{F}$ and $h_{z}=0.2E_{F}$. $\protect\alpha K_{F}=E_{F}$, $%
1/a_{s}K_{F}=-0.1$ for both (a) and (b). }
\label{gasless_property}
\end{figure}

\emph{Topological FF phase}: The transition from gapped FF phase to
topological FF phase is characterized by the quasiparticle excitation gap
that closes and reopens with increasing $h_{z}$ for a fixed $k_{z}$ and a
small $h_{x}$. During this transition, the minimum of the band gap $E_{g}$
occurs at $k_{x}=k_{y}=0$, and the gap closes when
\begin{equation}
(h_{x}+\alpha Q_{y}/2)^{2}+h_{z}^{2}=(\hbar ^{2}k_{z}^{2}/2m-\mu
)^{2}+\Delta _{0}^{2},  \label{zero_pts}
\end{equation}%
which determines the position $\mathbf{k}_{W}=\left( 0,0,k_{c}\right) $ of
the Weyl nodes. Such gap close and reopen is shown in Fig.~\ref{TS_property}%
(a) for $k_{z}=0$. During this transition, the order parameter $\Delta _{0}$
is still finite even the bulk gap is closed. The finite $Q_{y}$ indicates
the FF superfluid. In Fig.~\ref{TS_property}(b), we plot the Weyl nodes
determined by Eq. (\ref{zero_pts}) in the $(k_{z},h_{z})$ plane. For a fixed
$h_{z}$ and $h_{x}$, the superfluid is topological when there exists Weyl
nodes at certain $k_{z}$. There are two topological regions: $h_{z}>\sqrt{%
\mu ^{2}+\Delta _{0}^{2}-(h_{x}+\alpha Q_{y}/2)^{2}}$ with two Weyl nodes
and $\sqrt{\Delta _{0}^{2}-(h_{x}+\alpha Q_{y}/2)^{2}}<h_{z}<\sqrt{\mu
^{2}+\Delta _{0}^{2}-(h_{x}+\alpha Q_{y}/2)^{2}}$ with four Weyl nodes. Both
critical values for $h_{z}$ decrease with increasing $h_{x}$. To confirm
such excitations are Weyl fermions, we examine the energy dispersions around
these node points and find they are linear but with different slopes along
different directions (see Fig.~\ref{gasless_property}(a)). While the
different slopes between $k_{z}$ and in-plane directions are due to the 2D
SO coupling, the difference between $k_{x}$ and $k_{y}$ directions is caused
by the finite momentum $Q_{y}$ of the Cooper pairs induced by $h_{x}$. From
Eq. (\ref{zero_pts}), we see the properties of Weyl fermions, such as
position ($k_{c}$), number (2 or 4), and their creation and annihilation,
can be tuned by varying $h_{z}$ and $h_{x}$. Note that the anisotropy of
Weyl fermions can also be tuned by the external magnetic field in solid
materials (e.g. Pyrochlore Iridates~\cite{Aji2012prb}). In addition, such
Weyl fermions appear in pairs with opposite topological charges $N_{c}=\pm 1$
\cite{Note}.

Weyl nodes can also be regarded as quantum Hall transition points in the
momentum space parameterized by $k_{z}$. In the topological FF phase,
because the quasiparticle excitations are gapped (except at the Weyl nodes)
in the 2D plane with a fixed $k_{z}$, we can calculate the Chern number for
the hole branch for each $k_{z}$ plane
\begin{equation}
C\left( k_{z}\right) =\frac{1}{2\pi }\sum_{n}\int dk_{x}dk_{y}\Omega
^{n}(k_{x},k_{y}),  \label{Chern_equ}
\end{equation}%
where $n$ is the index for hole branches, and the Berry curvature \cite%
{XiaoRMP}
\begin{equation}
\Omega ^{n}=i\sum_{n^{\prime }\neq n}\left[ \frac{\langle n|\partial
_{k_{x}}H_{B}|n^{\prime }\rangle \langle n^{\prime }|\partial
_{k_{y}}H_{B}|n\rangle -(k_{x}\leftrightarrow k_{y})}{(E_{n\mathbf{k}%
}-E_{n^{\prime }\mathbf{k}})^{2}}\right] ,
\end{equation}%
and $n^{\prime }$, which is not equal to $n$, runs over the eigenstates of $%
H_{B}$. For the topological FF phase with two Weyl nodes, we find that $C=1$
when $|k_{z}|<k_{c}$ and $C=0$ when $|k_{z}|>k_{c}$ (see Fig.~\ref%
{TS_property} (c)). For the topological FF phase with four Weyl nodes, $C=1$
when $k$ lies between two nodes in the positive or negative $k_{z}$, and $C=0
$ otherwise. The Fermi surface lies at zero energy, which is composed of
separated Weyl nodes. It is expected that there exist chiral edge states~%
\cite{ZhongFang2011prl,Wan2011prb} in the presence of real space confinement
along $x$ or $y$ direction for a fixed $k_{z}$ with nonzero Chern number.
For instance, with the confinement along the $x$ direction, there are two
chiral edge states whose spectrum intersect at $k_{y}=0$ and $E=0$ ($k_{z}$
is already taken as a parameter). The zero energy Fermi surface now becomes
a line (i.e., Fermi arc) in the $(k_{y}$, $k_{z})$ plane that connects two
Weyl nodes (Fig.~\ref{TS_property}(d)).

\begin{figure}[t]
\includegraphics[width=1.7in]{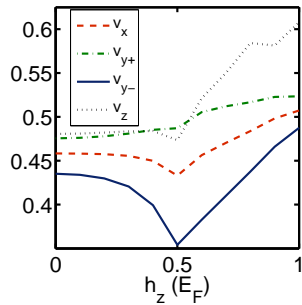}
\caption{(Color online) Sound speeds as a function of $h_{z}$. Dashed red
and dotted black lines represent the speed along $x$ and $z$ directions.
Solid blue and dashed-dot green lines represent sound speeds along negative $%
y$ and positive $y$ directions, respectively. The unit of sound speed is $%
v_{F}=\hbar K_{F}/m$. Here $h_{x}=0.2E_{F}$, $\protect\alpha K_{F}=E_{F}$,
and $1/a_{s}K_{F}=-0.1$.}
\label{sound}
\end{figure}

\emph{Gapless topological FF phase}: Topological FF state is gapless by
itself, but the gapless points occur only when particle and hole branches
touch. With increasing $h_{x}$, the quasiparticle excitation spectrum
becomes more asymmetric along the $k_{y}$ direction. Above certain $h_{x}$,
certain part of the particle branch of the spectrum may fall below the zero
energy, leading to gapless FF phase with $E_{g}<0$. Note that the band
minimum of particle branch and maximum of hole branch do not occur at the
same $\mathbf{k}$ in this phase. The gapless FF phase can be either
non-topological or topological, depending on whether the particle and hole
branches touch at certain $\mathbf{k}_{W}=\left( 0,0,k_{c}\right) $ with the
linear dispersion. Fig.~\ref{phase}(b) displays the zero energy contours in
the $(k_{y},k_{z})$ plane with $k_{x}=0$ for different states. The
topological FF phase has separate zero points along the $k_{z}$ axis, while
other phases have closed loops. For the gapless FF phase, the closed lines
are connected. For the gapless topological FF phase, the particle and hole
branch can still touch at certain $\mathbf{k}_{W}$. Instead of a single
point, the zero energy contour now cuts close loops in the $(k_{x},k_{y})$
plane and these loops are connected at $\mathbf{k}_{W}$. Away from $\mathbf{k%
}_{W}$ along the $k_{z}$ direction, particle-hole branches do not touch, and
the loops in the $(k_{x},k_{y})$ plane are disconnected, which can be
clearly seen from the bulk spectrum in Fig. \ref{gasless_property}(b). In
the $(k_{y},k_{z})$ plane with $k_{x}=0$, it forms two connected loop
structures around $\mathbf{k}_{W}$, as shown in Fig.~\ref{phase}(b). To
confirm the topological nontrivial feature, we calculate the Chern number of
the two hole bands and find it is one between two Weyl nodes and zero
otherwise. Fig.~\ref{gasless_property} (b) shows the edge state (red line)
in the gapless topological FF phase.

\emph{Sound speed of FF superfluids:} The anisotropic Weyl fermions in the
topological FF phase have two characteristic properties: (i) the anisotropic
energy spectrum along all three directions. Furthermore, the spectrum is
different even for $\pm k_{y}$ direction, which is special for the FF type
of superfluids. (ii) The energy gap closes above a critical $h_{z}$ for the
topological phase transition. In experiments, the finite momentum pairing of
the Cooper pairs may be measured using noise-correlation imaging~\cite%
{Greiner2005PRL} or momentum-resolved radio-frequency spectroscopy~\cite%
{Fu2013PRA}. Here we propose the above two characteristic properties of Weyl
fermions should manifest in the sound speeds of the underlying FF fermionic
superfluids, which can be probed by observing the propagation of a localized
density perturbation created by a laser beam, as demonstrated in previous
experiments \cite{Ketterle1997PRL,Joseph2007PRL}. The sound speed can be
obtained by calculating Gaussian fluctuations around the saddle point $%
(\Delta _{0},Q_{y})$ of the thermodynamical potential. {Specifically, the
speed of sound along $\eta $ direction is defined as $v_{\eta }=|$lim$%
_{q\rightarrow 0}\partial \omega (q)/\partial q_{\eta }|$, where $\omega (q)$
is bosonic gapless collective excitation spectrum \cite{Melo1997PRB} (see
supplementary information). }Around the Weyl nodes, where the quasiparticle
gap closes, we expect a sharp change of the sound speed because of the
strong fluctuations around the phase transition points. In Fig.~\ref{sound},
we see the speeds of sound {$v$}$_{i}$ are anisotropic along all three
different directions, indicating anisotropic quasiparticle spectrum. More
interestingly, the sound speeds along the positive and negative $y$
directions are also different, indicating asymmetric spectrum along the $y$
direction due to the finite momentum Cooper pairing. $v_{i}$ has the minimum
at the topological phase transition boundary (see Fig.~\ref{phase}(a)) to
the topological FF phase with Weyl fermions. Therefore, we conclude that
such anisotropic speeds of sound and the minimum located at the phase
transition boundary provide strong evidence of Weyl fermions in a FF
superfluid. In experiments, we consider a typical parameter with $^{40}$K
atoms and density $n=5\times 10^{12}\text{cm}^{-3}$, yielding the Fermi
energy $E_{F}=h\times 3.5$ KHz and Fermi velocity $v_{F}=8.3$ mm/s. The SO
coupling and Zeeman field can be created by Raman coupling between atomic
hyperfine states. The strength of the Zeeman field may be tuned by the
detuning and intensity of Raman lasers \cite{Lin2011Nature,Jing2012PRL,
Zwierlen2012PRL,PanJian2012PRL, Qu2013PRA,Spilman2013PRL,Zhang2010}.

\textit{Acknowledgement:} We would like to thank Xuele Liu, Li Mao, Lin
Dong, Ying Hu for helpful discussion and critical comments. This work is
supported by ARO (W911NF-12-1-0334), AFOSR (FA9550-13-1-0045), and NSF-PHY
(1249293).

\begin{widetext}
\section{Supplementary Materials}
In the main text we propose to detect the topological FF state by measuring
the anisotropic sound speeds, which show the signatures of finite momentum
pairings as well as topological phase transition. Here we present the
calculation details.

In quantum field theory, the partition function can be written as $Z=\text{Tr%
}(e^{-\beta H})=\int D(\bar{\psi},\psi )e^{-S_{eff}[\bar{\psi},\psi ]}$ with
$\beta =1/k_{B}T$. The effective action is
\begin{equation}
S_{eff}[\bar{\psi},\psi ]=\int_{0}^{\beta }d\tau \left( \int d\mathbf{r}%
\sum_{\sigma }\bar{\psi}_{\sigma }(\mathbf{r},\tau )\partial _{\tau }\psi
_{\sigma }(\mathbf{r},\tau )+H(\bar{\psi},\psi )\right) ,
\end{equation}%
where $\int d\tau $ is an integral over the imaginary time $\tau $ and $H(%
\bar{\psi},\psi )$ is obtained by replacing $\hat{\Psi}_{\sigma }^{\dagger }$
and $\hat{\Psi}_{\sigma }$ with Grassman field number $\bar{\psi}_{\sigma }$
and $\psi _{\sigma }$. We can transform the quartic interaction term to
quadratic one by Hubbard-Stratonovich transformation, where the order
parameter $\Delta (\mathbf{r},\tau )$ is defined. By integrating out fermion
fields, the partition function becomes $Z=\int D(\bar{\Delta},\Delta
)e^{-S_{eff}[\bar{\Delta},\Delta ]}$, where the effective action can be
written as
\begin{equation}
S_{eff}[\bar{\Delta},\Delta ]=\int_{0}^{\beta }d\tau \int d\mathbf{r}(\frac{%
|\Delta |^{2}}{U})-\frac{1}{2}\ln \det {G}^{-1}.
\end{equation}%
Here the inverse single particle Green function $G^{-1}=-\partial _{\tau
}-H_{B}$ in the Nambu-Gor'kov representation with $4\times 4$ Bogoliubov-de
Gennes (BdG) Hamiltonian
\begin{equation}
H_{B}=\left(
\begin{array}{cc}
H_{s}(\hat{\mathbf{p}}) & \Delta (\mathbf{r},\tau ) \\
\Delta (\mathbf{r},\tau ) & -\sigma _{y}H_{s}(\hat{\mathbf{p}})^{\ast }\sigma _{y}
\end{array}%
\right) .
\end{equation}%
Assume that a mean-field solution has the FF form $\Delta (\mathbf{r},\tau
)_{0}=e^{iQ_{y}y}\Delta _{0}$ with the space independent $\Delta _{0}$ due
to the in-plane Zeeman field that deforms the Fermi surface along the
\textit{y} direction, leading to finite momentum pairings along that
direction~\cite{YongArx13Sub}. This form of $\Delta (\mathbf{r},\tau )$ yields
the thermodynamical potential in the main text by Fourier transformation and
summing the Matsubara frequency.

To determine the collective modes, we explore Gaussian fluctuations around
the saddle point $\Delta (\mathbf{r},\tau )_{0}$ by separating the pairing
field as $\Delta (\mathbf{r})=\Delta (\mathbf{r},\tau )_{0}+A(\mathbf{r}%
,\tau )$ with fluctuation field $A$. Expanding $S_{eff}$ to the second
order, we obtain $S_{eff}=S_{eff}[\Delta _{0}]+\frac{1}{2}\sum_{q}\eta
^{\dagger }M\eta $. Here $\eta ^{\dagger }=(%
\begin{array}{cc}
A^{\ast }(q) & A(-q)%
\end{array}%
)$ with Fourier transformation $A(q)$ of $A(\mathbf{r},\tau )$. $q=(\mathbf{q%
},iq_{m})$ with $q_{m}=i2m\pi /\beta $. The $2\times 2$ inverse fluctuation
propagator reads
\begin{eqnarray}
M_{11}(q) &=&M_{22}(-q) \\
&=&\frac{1}{U}+\frac{1}{2}\sum_{k}\text{tr}\left[ \mathcal{G}_{0}(q+k)_{11}%
\mathcal{G}_{0}(k)_{22}\right] ,  \notag \\
M_{12}(q) &=&\frac{1}{2}\sum_{k}\text{tr}\left[ \mathcal{G}_{0}(q+k)_{12}%
\mathcal{G}_{0}(k)_{12}\right] , \\
M_{21}(q) &=&\frac{1}{2}\sum_{k}\text{tr}\left[ \mathcal{G}_{0}(q+k)_{21}%
\mathcal{G}_{0}(k)_{21}\right] ,
\end{eqnarray}%
where $\mathcal{G}_{0}(k)$ is the Fourier transformation of $G(\mathbf{r}%
,\tau )$; $\mathcal{G}_{0}(k)_{ij}$ indicates the $2\times 2$ block matrix
of $\mathcal{G}_{0}(k)$. $\sum_{k}$ represents summation over Matsubara
frequency and momentum $\mathbf{k}$. It is important to note that since the
analytical expression of $\mathcal{G}_{0}(k)$ cannot be obtained, we
calculate the summation over Matsubara frequency numerically.

The dispersion $\omega (\mathbf{q})$ can be determined by $\text{det}%
[M[\omega (\mathbf{q}),\mathbf{q}]]=0$ where $i\omega _{n}$ has been taken
analytically to $\omega (\mathbf{q})+i0^{+}$ at zero temperatures. To
decouple the amplitude and phase fluctuations, we define $A(x)=\rho
(x)+i\delta \theta (x)$, where $\rho (x)$ and $\delta \theta (x)$ are all
real. Now the second order effective action is
\begin{equation}
S_{eff}^{(2)}=\frac{1}{2}\sum_{q}\left(
\begin{array}{cc}
\rho (-q) & \delta \theta (-q)%
\end{array}%
\right) N(q)
\left(
\begin{array}{c}
\rho (q) \\
\delta \theta (q)%
\end{array}%
\right)=
\frac{1}{2}\sum_{q}\left(
\begin{array}{cc}
\rho (-q) & \delta \theta (-q)%
\end{array}%
\right)
\left(
\begin{array}{cc}
N_{11} & N_{12} \\
N_{21} & N_{22} \\
\end{array}%
\right) \left(
\begin{array}{c}
\rho (q) \\
\delta \theta (q)%
\end{array}%
\right) ,
\end{equation}%
where
\begin{eqnarray}
N_{11} &=&M_{11}(q)+M_{22}(q)+M_{12}(q)+M_{21}(q), \\
N_{12} &=&iM_{11}(q)-iM_{12}(q)+iM_{21}(q)-iM_{22}(q), \\
N_{21} &=&-iM_{11}(q)-iM_{12}(q)+iM_{21}(q)+iM_{22}(q), \\
N_{22} &=&M_{11}(q)+M_{22}(q)-M_{12}(q)-M_{21}(q).
\end{eqnarray}

To study gapless low energy excitations, we make a small $\mathbf{q}$ and $%
\omega $ expansion~\cite{Melo1997PRBSub} using
\begin{eqnarray}
N_{11} &=&A^{11}+\sum_{\sigma =x,y,z}q_{\sigma }^{2}B_{\sigma }^{11}-\omega
^{2}C^{11}+\ldots , \\
N_{22} &=&\sum_{\sigma =x,y,z}q_{\sigma }^{2}B_{\sigma }^{22}-\omega
^{2}C^{22}+\ldots , \\
N_{12} &=&N_{21}^{\ast }=-i\omega C^{12}+iq_{y}D_{y}^{12}+\ldots .
\end{eqnarray}

Here $D_{y}^{12}$ comes from the nonzero momenta of Cooper pairs along the $y
$ direction. The equation $\det (N)=N_{11}N_{22}-|N_{12}|^{2}=0$ leads to
\begin{equation}
\left[ A^{11}C^{22}+(C^{12})^{2}\right] \omega ^{2}-2\omega
q_{y}C^{12}D_{y}^{12}+q_{y}^{2}(D_{y}^{12})^{2}-A^{11}\sum_{\sigma
}q_{\sigma }^{2}B_{\sigma }^{22}=0.
\end{equation}%
For a special example of a uniform superfluid phase, $D_{y}^{12}=0$, $%
B_{x}^{22}=B_{y}^{22}$, and the collective mode dispersion is
\begin{equation}
\omega (\mathbf{q})=\sqrt{\frac{A^{11}\left[
B_{x}^{22}(q_{x}^{2}+q_{y}^{2})+B_{z}^{22}q_{z}^{2}\right] }
{A^{11}C^{22}+(C^{12})^{2}}}.
\end{equation}%
The sound speed along the $\sigma$ direction with $\sigma=x,y,z$  is
\begin{equation}
s_{\sigma }=\sqrt{\frac{A^{11}B_{\sigma }^{22}}{A^{11}C^{22}+(C^{12})^{2}}},
\end{equation}%
which is anisotropic between $z$ and in-plane directions because of the 2D
SO coupling. In the absence of spin-orbit coupling, $%
B_{x}^{22}=B_{y}^{22}=B_{z}^{22}$ and the formula reduces to the well-known
result~\cite{Melo1997PRBSub}. For a general case of a FF superfluid, $%
D_{y}^{12}\neq 0$, the dispersion along different directions are
\begin{eqnarray}
\omega (q_{x},q_{y}=0,q_{z}=0) &=&v_{x}|q_{x}|, \\
\omega (q_{x}=0,q_{y}=0,q_{z}) &=&v_{z}|q_{z}|, \\
\omega (q_{x}=0,q_{y}>0,q_{z}=0) &=&v_{y+}q_{y}, \\
\omega (q_{x}=0,q_{y}<0,q_{z}=0) &=&-v_{y-}q_{y},
\end{eqnarray}%
where the speeds of sound are
\begin{eqnarray}
v_{\mu } &=&\sqrt{\frac{A^{11}B_{\mu }^{22}}{A^{11}C^{22}+(C^{12})^{2}}}, \\
v_{y+} &=&(-E+\sqrt{E^{2}+4F})/2, \\
v_{y-} &=&(E+\sqrt{E^{2}+4F})/2,
\end{eqnarray}%
with $\mu =x,z$, and
\begin{eqnarray}
E &=&-\frac{2C^{12}D_{y}^{12}}{A^{11}C^{22}+(C^{12})^{2}}, \\
F &=&-\frac{(D_{y}^{12})^{2}-A^{11}B_{y}^{22}}{A^{11}C^{22}+(C^{12})^{2}}.
\end{eqnarray}%
We see that the difference of the sound speeds between $x$ and $y$
directions originates from $D_{y}^{12}$ due to finite momentum pairings,
which also leads to the speed's difference, equal to $E$ proportional to $D_y^{12}$,
along the positive and negative $y$ directions.

\end{widetext}


\begin{thebibliography}{99}
\bibitem{Note} The topological charge is defined as $N_{c}=\frac{1}{24\pi
^{2}}e_{\mu \nu \lambda \gamma }\text{tr}\int_{\sum_{a}}dS^{\gamma }%
\mathscr{G}\partial _{p_{\mu }}\mathscr{G}^{-1}\mathscr{G}\partial _{p_{\nu
}}\mathscr{G}^{-1}\mathscr{G}\partial _{p_{\lambda }}\mathscr{G}^{-1}$~\cite%
{volovik}, where the Green function $\mathscr{G}^{-1}=ik_{0}-H_{B}(\mathbf{k}%
)$, the integral is over the 3D surface $\sum_{c}$ embracing the isolated
Weyl node $\mathbf{k}_{W}=(0,0,k_{c})$, and $\text{tr}$ stands for the trace
in the spin space.
\end{thebibliography}
\end{document}